\documentclass[twocolumn,twocolappendix,lineno,trackchanges]{aastex631}
\shorttitle{A dichotomy in the outbursts of AM CVns}
\shortauthors{Rivera Sandoval et al.}
\graphicspath{{./}{figures/}}

\begin{document}

\title{The Fast Evolving, Tremendous and Blue Superoutburst in ASASSN-21au Reveals a Dichotomy in the Outbursts of Long-period AM~CVns}

\correspondingauthor{L. E. Rivera Sandoval}
\email{liliana.riverasandoval@utrgv.edu}

\author{L. E. Rivera Sandoval}
\author{C. O. Heinke}
\affiliation{University of Alberta, Department of Physics, CCIS 4-183, T6G 2E1, Edmonton, AB, Canada}

\author{J. M. Hameury}
\affiliation{Observatoire astronomique de Strasbourg, Université de Strasbourg, CNRS UMR 7550, 67000 Strasbourg, France}

\author{Y. Cavecchi}
\affiliation{Universidad Nacional Aut\'onoma de M\'exico, Instituto de Astronom\'ia, Ciudad Universitaria, CDMX 04510, Mexico}

\author{T. Vanmunster}
\affiliation{CBA Belgium Observatory \& CBA Extremadura Observatory, Walhostraat 1a, B-3401 Landen, Belgium.}

\author{T. Tordai}
\affiliation{Polaris Observatory, Hungarian Astronomical Association, Laborc u. 2/c, 1037 Budapest, Hungary}

\author{F. Romanov}
\affiliation{AAVSO observer}



\begin{abstract}

 ASASSN-21au is an ultracompact accreting white dwarf binary (AM~CVn) with a period of $\sim58$~min. Using multiwavelength observations of the system, we discovered a dichotomy in the behavior of outbursts in AM~CVns. The binary showed an initial brightness increase which lasted for at least 82 days, followed by an additional increase which lasted 2 weeks. Afterwards ASASSN-21au went into superoutburst with a total duration of 19 days, showing an amplitude with respect to quiescence of $\sim7.5$~mags in $g$, with a precursor and an echo outburst. A correlation between X-rays, UV and optical was identified for the first time in an AM~CVn during this stage. The color evolution of ASASSN-21au indicates that during the superoutburst the dominant component was the accretion disk. The short duration, large amplitude and color evolution of the superoutburst agree with expectations from the disk instability model. 
These characteristics are opposite to the ones observed in SDSS~J080710+485259 and SDSS~J113732+405458, which have periods of $\sim53$~min and $\sim60$~min, respectively. The initially slow brightness increase in the light curve of ASASSN-21au and the behavior after the superoutburst favors a scenario in which changes in the mass-transfer rate led to disk instabilities, while the outburst mechanism of SDSS~J080710+485259 and SDSS~J113732+405458 has been attributed to enhanced mass-transfer alone. Further observations are needed to understand the origin of this dichotomy.

\end{abstract}


\keywords{Stars: individual (ASASSN-21au); AM Canum Venaticorum stars; White dwarf stars; Compact binary stars; Stellar accretion disks; Dwarf novae; Hydrogen deficient stars; Interacting binary stars; Cataclysmic variable stars; Transient detection; Stellar accretion}


\section{Introduction} \label{sec:intro}
\label{intro}

AM~CVns are a 
relatively poorly studied
class of accreting white dwarf (WD) binaries. Having typical orbital periods (P$_{\rm orb}$) between 5 and 70 min \citep[e.g.][]{2020Green,2018ramsay}, AM~CVns 
include 
the accreting binaries with the shortest orbits known so far, which makes them important 
sources of
low frequency gravitational waves 
which can
be detected by LISA \citep[][ and references therein]{2018Breivik,2021Liu}.\looseness=-10
 
 AM~CVns share several similarities with 
 the cataclysmic variables (CVs), especially
 dwarf novae (DNe).
 However, unlike CVs where the donor is hydrogen-rich, AM CVns are hydrogen-poor. The latter can be produced through three principal channels, depending on the initial state of the donor when mass transfer starts: the donor can be a fully degenerate white dwarf, either He or C/O 
 \citep[e.g.][]{1975Pringle,1979Tutukov}, a semi-degenerate helium core-burning star 
 \citep{1986Savonije}, or a subgiant \citep[leading to the exposure of the semi-degenerate hydrogen-deficient core;][]{1985Tutukov,1986Nelson}. Depending on which evolutionary channel is followed, the chemical composition of the donor will change \citep[e.g.][]{2010Nelemans}.
 The orbital periods of CVs are also much longer,
 with typical values of 85~min to 10~hrs.

As occurs in the case of 
CVs, the evolution of AM~CVns after reaching the period minimum \citep[e.g.][]{1991Iben,2008yungelson} dictates that the mass-transfer rate from the companion reduces as the binary evolves. 
The degenerate nature of the donor star 
causes the binary's orbit to become wider with time
\citep[e.g.][]{1985Tutukov}. These binaries also 
become progressively 
fainter with time, which makes them difficult to detect when 
in quiescence. 
However, for the AM~CVns that show a transient behavior (like DNe in regular CVs), their brightness increases by several magnitudes during DN-like outbursts, making them detectable to all sky optical surveys \citep[e.g.][]{2013levitan,2021vanroestel}. 
Most AM~CVns known so far have been identified in this way.

According to the standard disk instability model \citep[see review by][and references therein]{2020Hameuryreview},
the presence of outbursts in AM~CVns depends on the value of the mass-transfer rate. Observationally 
a dividing line
occurs at periods
around
P$_{\rm orb} = 20$~min. For shorter periods, the high mass-transfer rate ensures the disks are hot and stable, hence no outbursts are expected, while at longer periods the disk builds up to intermittent outbursts. However, unlike the case of CVs, for AM~CVns the mass-transfer rate at relatively long orbital periods is expected to be so low 
that the accretion disk becomes cold and stable again, so outbursts are not expected to be observed either. The lower orbital period limit for that phenomenon is
not really clear, either theoretically (due to the several free parameters in the models) or observationally.

Further complications have been recently added by observations of long-period AM~CVns in outbursts \citep{2020RS08,2021RS}, which suggest that enhanced mass-transfer 
also plays a very important role as
an outburst triggering mechanism.

In this paper we present multiwavelength  observations of a recently discovered AM~CVn star, known as ASASSN-21au, which was identified spectroscopically during outburst and which had a variable superhump period (a period in which the disk precesses, close to the orbital period; \citealt{Patterson05}) between 
$57-60$~min \citep[][vsnet-alert 25369]{2021Isogai}\footnote{We have chosen to use an average superhump period of $58.4$~min as obtained from a Lomb-Scargle \citep{Lomb1976,Scargle1982} analysis.}.
The observations of ASASSN-21au here discussed demonstrate that there is a dichotomy in the behavior of outbursts in long-period AM~CVns.

\section{Observations and Data Analysis}

\subsection{Swift X-ray and UV Observations}

We have obtained X-ray and UV data of ASASSN-21au during its first superoutburst and post-outburst cooling phase with the XRT and UVOT instruments on board the Neil Gehrels Swift Observatory (\emph{Swift}). Data were taken from 2021-02-17 to 2021-06-11 with a total exposure time of $12$~ks distributed in 14 observations.  

ASASSN-21au was detected in X-rays in 9 of the 
observations, with upper 
limits on the remaining measurements, mainly due to the short exposure times. The XRT data analysis was performed using XSPEC \citep[][Version 12.11]{199arnaud}. Count rates were obtained in the X-ray energy range 0.3-10 keV for each of the individual observations. For the spectral analysis we used an absorbed power-law model ({\it tbabs*powerlaw}) fixing $N_h$ to the Galactic value of $2.87\times10^{20}$~cm$^{-2}$ toward the position of the source. In order to obtain meaningful values for the spectral fits we divided the observations in 2 groups: those obtained during the plateau phase of the superoutburst and those taken during the post-outburst cooling phase where the binary was detected. 
For each group we binned the data considering at least 5 counts per bin and used Cash-statistics \citep{1979cash} due to the low number of counts acquired. \looseness=-10

For each individual \emph{Swift} observation, UVOT data were obtained in the 6 available filters: $UVW2, UVM2, UVW1, U, B, V$ which in total cover the wavelength range from $1600$ to $8000$~\AA. Data were analysed with the suggested tools in the \emph{Swift} threads\footnote{\url{https://www.swift.ac.uk/analysis/uvot/}}. For the  
background
subtraction we used a circular region of $5\arcsec$ radius around the target, and a circular region with a radius of $25\arcsec$ for the background determination. 
Photometry was calibrated to the AB system. \looseness=-10

\subsection{Optical and Additional UV Observations}

Data in the clear and V filters were obtained by members of the AAVSO \citep[][]{AAVSO_cit} and VSNET with the first measurement starting on 	
2021-02-12 and the last one taken on 	
2021-04-15. Several epochs with cadences of 1 min per observation were obtained. We have aligned the data from the different observers into a common frame by using the first dataset of observations as reference, which is consistent with the results from the \emph{Swift} $V$ filter. 

For comparison purposes we have also made use of public observations from the 
ATLAS project in the cyan and orange bands \citep{2018Tonry} and from ZTF \citep[][]{2019masci} in the $g$ and $r$ bands. Only data with good quality flags have been used for the analysis. Additionally, in the case of the ZTF data, we excluded points with airmass $ >1.8$ since the differential chromatic refraction that produces color biases dominates above that value \citep{2019masci}.
We also used data from ASAS-SN \citep{2014Shapee, 2017Kochanek} in the $g$ band, data from Pan-STARRS \citep{2016Chambers} in the $g, r, i$ and $z$ bands, from GALEX \citep{2011Bianchi} in the $NUV$, from Gaia EDR3 \citep{2020gaia}
and the Gaia  alerts\footnote{\url{http://gsaweb.ast.cam.ac.uk/alerts/alert/Gaia21cbs/}} in the $G$ band. Photometry from these surveys is here reported in the AB system. 

We also checked for indications of previous outburst activity from the binary in the ATLAS, ASAS-SN, ZTF,
 CSS \citep{2009Drake} and  
DASCH \citep{2009grindlay} databases, but no outbursts of similar amplitude to the one reported here were previously recorded.

\section{Results}

\subsection{The Multiwavelength Light Curve of ASASSN-21au}
\label{light_curve_sec}

\begin{figure}
	\includegraphics[width=1\columnwidth, trim=0cm 0cm 0cm 0cm]{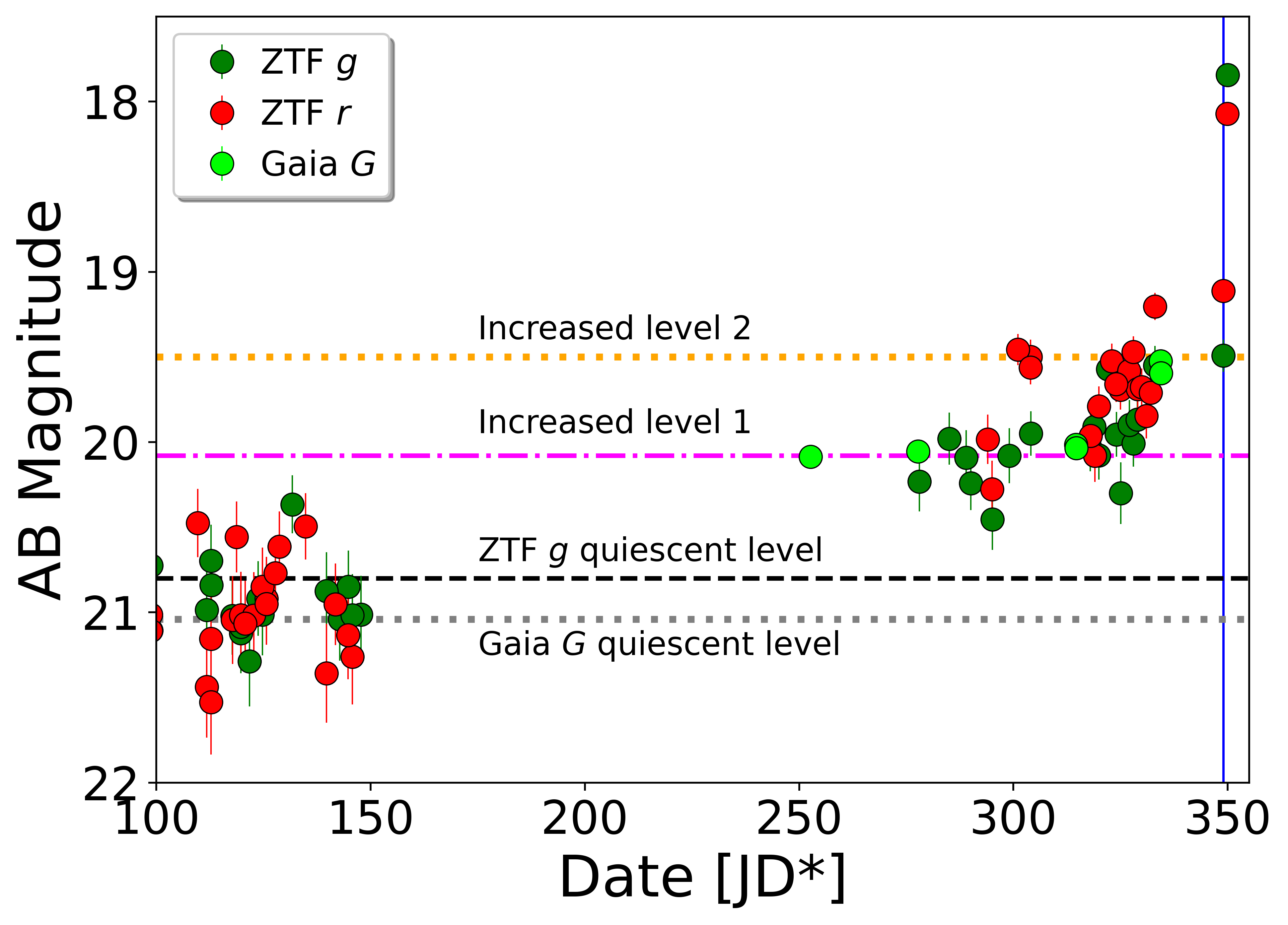}
    \caption{Optical light curve of ASASSN-21au during quiescence and the first indications of brightness increases. The blue vertical line denotes the beginning of the precursor of the superoutburst.}
    \label{fig:quiescence}
\end{figure}

\begin{figure}
	\includegraphics[width=1\columnwidth, trim=0.5cm 0.5cm 0cm 0cm]{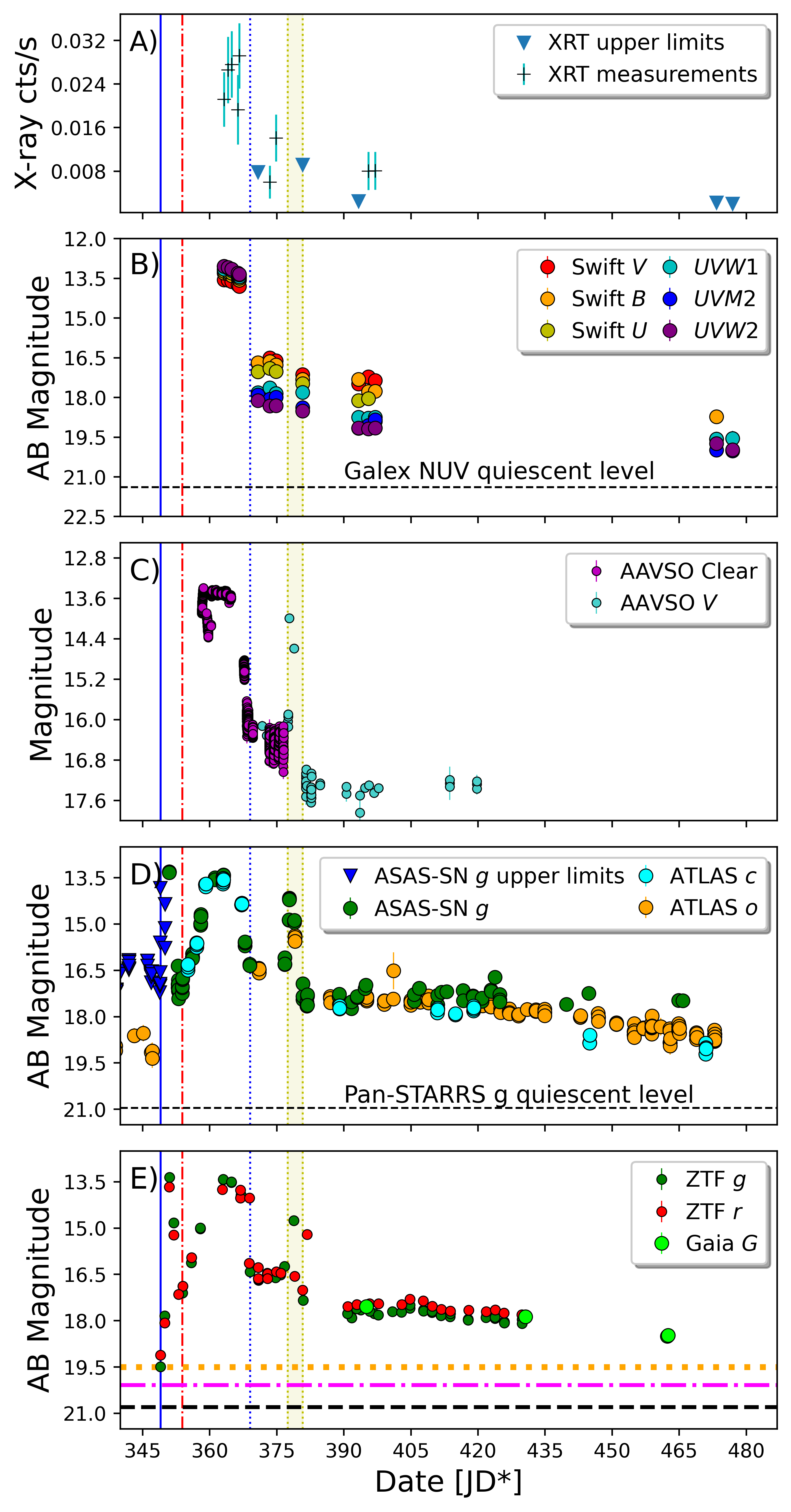}
    \caption{
    {\footnotesize 
Multiwavelength light curve of ASASSN-21au during superoutburst.
The beginning was determined using ZTF data and is marked with a blue solid line.
the blue dotted line indicates the end of the superoutburst and the beginning of the post-outburst cooling phase. The dashed red line marks the end of the precursor and the yellow band denotes the period where an echo outburst occurred between JD* 378 and 381.
    A):  XRT lightcurve in the energy range $0.3-10$~keV. 
    X-ray detections during the post-outburst cooling phase are signatures of residual accretion. 
    B): Multiband UVOT observations. They are simultaneous with the X-ray data. 
    The GALEX NUV quiescent level is indicated with a black dashed line.
    C): Fast (1 min) cadence AAVSO data. Periodic oscillations due to superhumps were detected in the superoutburst. 
    D): Light curves with data from the ASAS-SN and ATLAS surveys. 
    E): Gaia and ZTF light curves. The beginning of the precursor was detected on JD*=349. The colors of the horizontal lines have the same meaning than those in Figure \ref{fig:quiescence}.
    }
    }
    \label{fig:LC}
\end{figure}

The multiwavelength light curves of ASASSN-21au are shown in Figure \ref{fig:quiescence} and Figure \ref{fig:LC}. The first brightness increase of ASASSN-21au was detected by Gaia on 2020-10-30 ($JD*=252.7$, where $JD*=JD-2458900$) when the binary was 1 mag above (marked as increased level 1) its original $G$ quiescence level. On 2020-11-24 ($JD*=278$) ZTF $g$ detected the binary at the same increased level~1 (corresponding to $\sim0.8$~mags above its ZTF~$g$ quiescent level). Constraints due to the Sun prevented a good coverage with ZTF to observe the rise phase, but based on the current ZTF and Gaia data, a limit on the duration of the rise and the increased level~1 ($\tau_{L1}$) can be determined as $67$ days $<\tau_{L1}<172$ days. On $JD*=333$ the binary reached a second increased brightness level (marked with an orange line in Figures \ref{fig:quiescence} and \ref{fig:LC}), which was 1.5 mags above the original Gaia $G$ quiescent value. The system took approximately 2 additional weeks to reach that level and remained in such state for another 2 weeks. On 2021-02-03 ($JD*=349$) ASASSN-21au showed a sudden brightness increase due to the precursor of the superoutburst (denoted with a blue vertical line in Figures \ref{fig:quiescence} and \ref{fig:LC}).\looseness=-10

The first ASAS-SN solid detections of ASASSN-21au were
three points at $g\sim13$~mags on 2021-02-05 (JD*=351), followed by several other points 2-3 days later at $g\sim16.4-17.4$~mags, which indicate the fading of the precursor. Afterwards the rise to the plateau phase started. 
The maximum
of the precursor was similar to the 
maximum of the plateau 
($\sim7.5$ mags in $g$ above its ZTF quiescent level).
While precursors are common in DNe of the type SU~UMa and they have been recently confirmed in multiple AM~CVns \citep{2021PichardoM, DuffyTESSKLDRA2021}, they are relatively difficult to detect using ground-based images due to their short duration. But continuous coverage of AM~CVns' superoutbursts with space telescopes now suggest that precursors are a relatively common characteristic \citep{2021PichardoM}. In the case of the precursor of ASASSN-21au, it is also remarkable that its luminosity drops by 3 magnitudes right before the superoutburst begins, much more than in any other system \citep{2021PichardoM, DuffyTESSKLDRA2021}.\looseness=-10

\begin{figure}
	\includegraphics[width=1\columnwidth, trim=0.5cm 0cm 0.5cm 0cm]{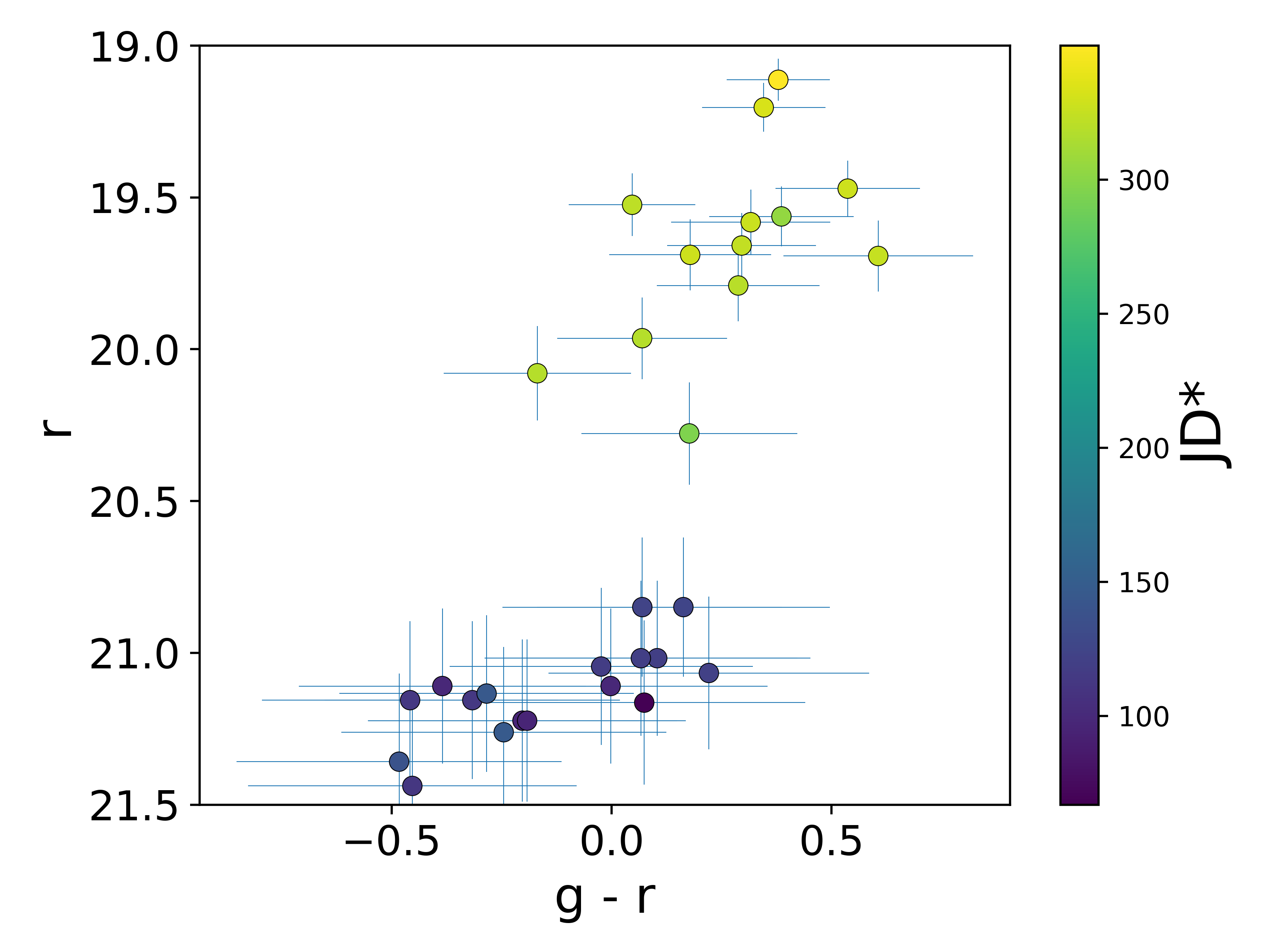}
    \caption{Color evolution for ASASSN-21au during quiescence and the increased levels 1 and 2 as shown in Figure \ref{fig:quiescence} using ZTF data. The binary is redder during the period of increased brightness.}
    \label{fig:colorZTF}
\end{figure}

In the bluest UV band ($UVW2$), there was a decrease 
of 3 orders of magnitude in flux between the plateau of the superoutburst and the post-outburst cooling phase around $JD*=373.3$, going from $1.60\times 10^{-13}$ erg s$^{-1}$ cm$^{-2}$ to $3.36\times 10^{-16}$ erg s$^{-1}$ cm$^{-2}$, while in the $V$ band 
the flux decreased by only
2 orders of magnitude, showing that the dominant emission components during superoubursts were the accretion disk, the boundary layer and the accreting WD. 

The superoutburst lasted for 15 days, excluding the precursor, and ended on 2021-02-23 (or $JD*=369$) as indicated by the data presented in Figure \ref{fig:LC}. This superoutburst was followed by an echo outburst, occurring more than a week after the superoutburst ended; these are frequent in SU~UMa systems that show only superoutbursts, and have been interpreted by \citet{2021hameury} as the manifestation of an increased mass-transfer rate, possibly as a result of the secondary irradiation by the heated accreting WD at the end of a superoutburst. This interpretation is strengthened by the fact that in ASASSN-21au, the system has not returned to full quiescence at the end of the superoutburst.

We now compare the behavior between the X-rays, optical and UV observations obtained by UVOT. As shown in other AM CVns, during the post-outburst cooling phase of ASASSN-21au there is X-ray emission indicating the presence of residual accretion. Interestingly, during 
the post-outburst cooling phase the spectrum was substantially redder than during the peak of the superoutburst. A possible explanation for such a behavior is that the inner parts of the disk may have evaporated in that cooling phase.

Figure \ref{fig:LC} also shows a correlation between the optical, UV and X-rays. During the optical and UV maxima, 
the X-ray emission also peaked, and during the optical and UV decay to the post-outburst cooling phase the X-ray flux decreased
 by a factor of 3. This behavior is different 
from that observed in the AM~CVns KL~Dra \citep{2012ramsay_sup} and SDSS~J141118+481257 \citep[henceforth SDSS~1411;][]{2019RS}, where an anticorrelation between the optical/UV and the X-rays during superoutburst was observed. The same anticorrelation has been observed in many CVs \citep[e.g.][]{2003wheatley, 2009Byckling, 2011Fertig}, where it has been explained as due to changes in the optical depth of the boundary layer. 
 
 According to the standard theory \citep{PattersonRaymond85a,PattersonRaymond85b}, increases in the X-ray flux with accretion rate are expected so long as the boundary layer remains optically thin; once the boundary layer becomes optically thick, the majority of the flux shifts to the far-UV and the X-ray flux drops.  
 Three nearby CVs, SS Cyg, U Gem and GW Lib, have shown a clear correlation between X-ray and optical/UV flux. SS Cyg showed this during the early, and late, parts of its outburst, with an anticorrelation at the highest optical fluxes (and, presumably, mass-transfer rates; \citealt{2003wheatley}). \citet{2011Fertig} summarize observations of 6 well-studied DN outbursts, in which 4 showed suppression of the X-rays with increased mass-transfer at some point. The peak X-ray luminosities of these CVs (in this paradigm, reflecting the critical accretion rate for the boundary layer to turn optically thick) mostly lie between $L_X=1\times10^{32}$ and $1\times10^{33}$ erg/s, though VW Hyi showed X-ray suppression from a peak of only $6\times10^{30}$ erg/s.  
 
 However, the critical accretion rate for the boundary layer turning optically thick may be different for He vs. H accretion. 
 The peak $L_X$ values for the two AM~CVn systems with well-studied outbursts where an anticorrelation has been observed have $5\times10^{30}$ erg/s (KL~Dra, \citealt{2012Ramsay}) using the Gaia distance of 948 pc \citep{Bailer-Jones21}, and $2\times 10^{31}$ erg/s (SDSS~1411, \citealt{2019RS}).
 These are substantially smaller than the typical values for H-accreting CVs, suggesting that the critical accretion rate is systematically lower for He accretion. Assuming that ASASSN-21au reaches a peak $L_X$ below the peak $L_X$ of the other two AM CVn systems (to account for its X-ray/optical correlation), its inferred distance should be below 400 pc.
 However, since we do not understand the origin of the large scatter in peak $L_X$ values we cannot be too confident of this estimate. It is also possible that the X-ray emitting region is still optically thin if the accretion disk does not extend to the accreting WD surface, but is, even at large accretion rates, truncated by, e.g., a strong magnetic field.

We do not see clear
changes in the X-ray spectral index of ASASSN-21au 
between
the plateau 
and
the post-outburst cooling phase, despite the changes in flux. Photon indexes of  $\Gamma=2.2\pm0.28$ and $\Gamma=2.08\pm0.56$ were determined for each phase, 
consistent within the errors.  

\subsection{ASASSN-21au and the Dichotomy in the Outburst Behavior of AM CVns}

\subsubsection{Outburst Duration: Disk Instabilities versus Enhanced Mass-Transfer}

The light curves of ASASSN-21au reveal several interesting characteristics. 
First, the initial scaled brightness increases up to day $JD*=349$ are slow, of low amplitude and with a so-called red color evolution (Figure \ref{fig:colorZTF} and \S \ref{color}), reaching ZTF $g-r > 0.5$. Those are the same characteristics observed in the outburst light curves of the AM~CVns SDSS~J080710+485259 (henceforth 
SDSS~0807) and SDSS~J113732+405458 \citep[henceforth SDSS~1137;][]{2020RS08, 2021RS, 2021Wong}, which have P$_{\rm orb}$ close to $53$ and $60$ min, respectively.

Second, on $JD*=349$ ASASSN-21au developed a short duration superoutburst, which contrasts with the longer than a year duration outbursts of SDSS~0807 and SDSS~1137, suggesting that there is more than one mechanism at work in the outbursts of long-period AM~CVns. On one hand a superoutburst lasting for 19 days (including the precursor), with a short rise time, and plateau phase followed by an abrupt cut-off, as observed in ASASSN-21au is fully consistent with expectations from the disk instability model \citep[DIM;][]{2015cannizzo, 2019cannizzo}.
On the other hand, outbursts lasting for more than a year cannot be explained by the DIM, for the simple reason that the viscous time \citep[$t_{\rm visc}$;][]{2002fkr}
is far too short.
The DIM predicts that, during the outburst plateau\footnote{Note that the plateau phase does not mean that there is no evolution, but that the emission evolves much slower than during the rise and decline phases.}, the disk is in a quasi-steady state and evolves exponentially on this timescale:

\begin{equation}
t_{\rm visc} = 6.43 \alpha^{-4/5} \dot M_{16}^{-3/10} M_1^{1/4} r_{10}^{11.4} \; \rm d,
\end{equation}

where $\alpha \sim 0.2$ is the Shakura-Sunyaev parameter in the hot state, $\dot M_{16}$ is the mass accretion rate in $10^{16}$~g~s$^{-1}$ units, $M_1$ the primary mass in solar units and $r_{10}$
is the outer disk radius in 10$^{10}$~cm, with $r_{10} \sim 1 - 1.5$ for an orbital period close to 1~hr.
$t_{\rm visc}$ is much smaller than the observed duration of the whole outburst. An increase of the mass-transfer due to irradiation of the secondary does increase the outburst duration, but under the DIM with enhanced mass-transfer for the systems with outbursts longer than a year one would need fine tuning to change the outbursts duration by almost two orders of magnitude. There are indications from ATLAS data that SDSS~1137's outburst lasted $\sim650$ days\footnote{ T. Kupfer and J. van Roestel, private communication.}. If the duration of SDSS~1137's outburst was in fact longer than previously reported \citep{2021RS,2021Wong}, the difference in duration with ASASSN-21au would be even more remarkable. 
Another interesting difference among these long-period AM~CVn systems is that the $U$ emission in SDSS~1137 seems to have been suppressed, as there was no evidence of an increase in that band \citep{2021RS}, while in the case of ASASSN-21au there is a clear increase in $U$ during superoutburst (Figure \ref{fig:LC}).

\begin{figure}
	\includegraphics[width=1.1\columnwidth, trim=0cm 0cm 0cm .5cm]{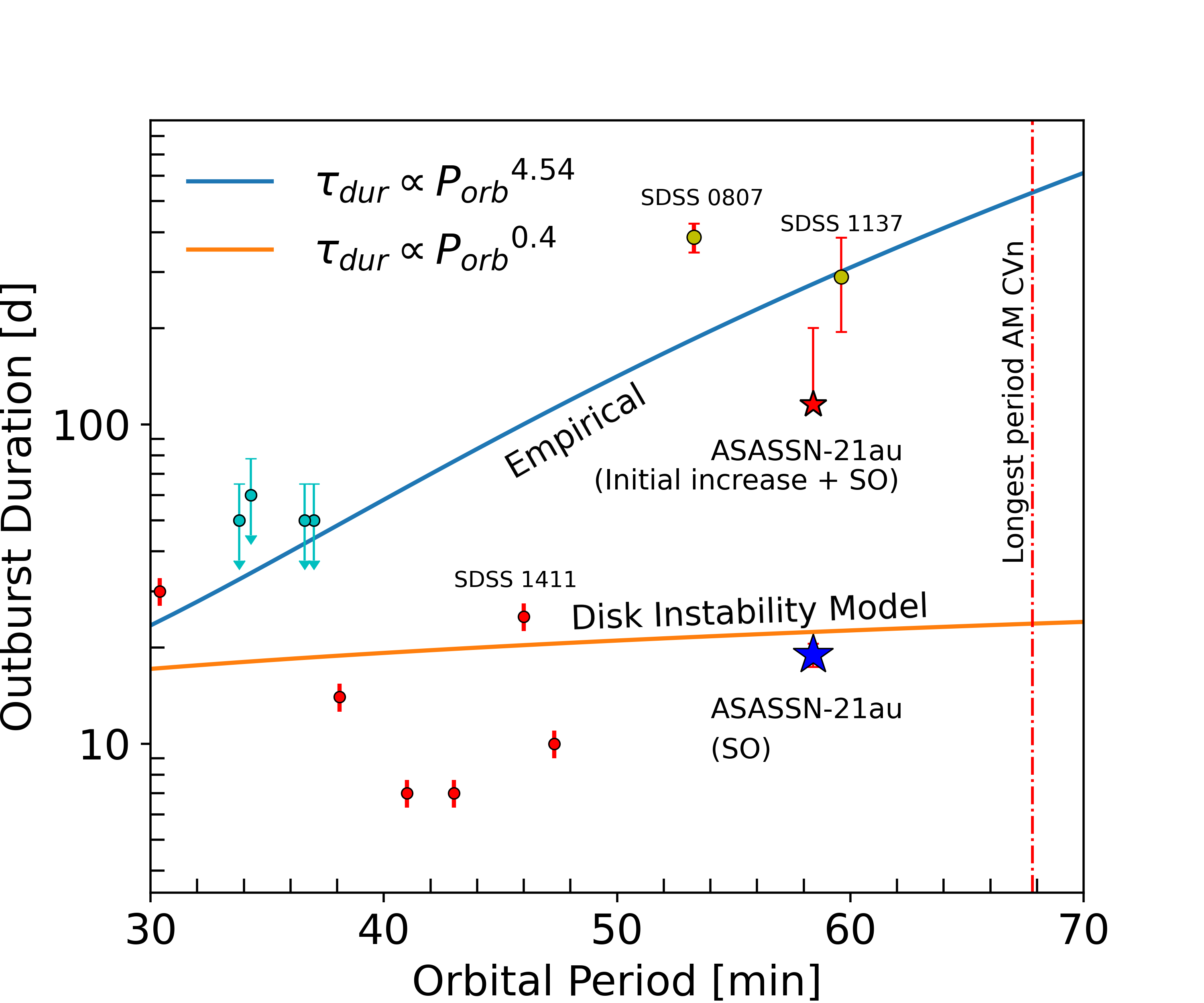}
    \caption{Orbital period (P$_{\rm orb}$) vs outburst duration ($\tau_{dur}$) for AM~CVns. The duration of the superoutburst (SO) of ASASSN-21au is indicated by a blue star. A star in red color indicates the duration of the superoutburst including the initial brightenings. Upper error for that red star symbol has been overestimated and it has been determined from the last quiescent ZTF point to the first Gaia detection with increased brightness. The long-period AM~CVn SDSS~0807 which had an outburst with a duration longer than a year is marked in yellow color. In the case of SDSS~1137 we have plotted the duration as reported in \citet{2021RS} but there are indications that the event lasted around 650 days. Red points are other AM~CVns as reported by \citet{2019cannizzo} and references therein. The cyan points are upper limits.
    The orange line indicates the relation between P$_{\rm orb}$ and $\tau_{dur}$ obtained from the DIM by \citet{2019cannizzo} and the blue line is the empirical relation derived by \citet{2015Levitan}. The duration of the SO in ASASSN-21au is consistent with expectations from the DIM, but it is several tens of times shorter than the one of SDSS~0807 and SDSS~1137 clearly indicating the existence of a dichotomy.
    The red dashed line indicates the period of the AM~CVn with the longest P$_{\rm orb}$ detected so far \citep{2020Green}. Up to date no outbursts have been detected from that binary. }
    \label{fig:relation}
\end{figure}  
 
 In Figure \ref{fig:relation} we compare the duration of the superoutburst of ASASSN-21au to the ones of other AM~CVns. We have also plotted the duration of the superoutburst of ASASSN-21au when we include the period of increased brightness previous to the superoutburst. This figure shows that there is a clear dichotomy in the outburst duration of long-period AM~CVns.
 Based on the behavior of ASASSN-21au up to day $JD*=349$, and the similarities with SDSS~0807 and SDSS~1137, it seems that the mechanism that caused the brightness increases in ASASSN-21au favors an enhanced mass-transfer scenario as well. However, the subsequent development of a DIM outburst in ASASSN-21au contrasts with the outburst evolution of SDSS~0807 and SDSS~1137. We stress that inclination effects do not explain such dichotomy given that eclipsing AM~CVns (ZTFJ0407-00, YZ~LMi, PTF1J1919+4815, Gaia14aae) show
blue superoutbursts of short duration and large amplitude \citep[e.g.][]{2021vanRoestel_eclipsing,DuffyTESSKLDRA2021, 2015campbell}.\looseness=-10
 
\subsubsection{Outburst Decay Times}

For comparison purposes of the decay times, we have fitted the time evolution of the magnitude with a straight line and determined its slope $n$ for ASASSN-21au, SDSS~0807 and SDSS~1137 during the plateau's decline and post-outburst cooling phase using the ZTF data for consistency (Table \ref{decays}). However, for the case of ASASSN-21au we also used the AAVSO clear data which had a better coverage in time for the decay from the plateau ($367<JD*<369$) and early times of the cooling phase ($369<JD*<376.5$, n=($21.22\pm0.84$)$\times10^{-3}$ mag/d). For this binary we also fitted the decay from the echo outburst ($377.8<JD*<381$) using the ASAS$-g$ observations and obtained $n=(1179.80\pm46.05)\times10^{-3}$ mag/d. The latter value is consistent with that obtained when fitting the decay part of the superoutburst using the AAVSO data in the clear filter (Table \ref{decays}). Note that in the case of SDSS~0807 there is substantial 
scatter in both ZTF bands ($r$ and $g$) and thus the fit is affected by that. 
From this analysis one sees that the decay of ASASSN-21au is $\sim3$ orders of magnitude steeper than in the other AM~CVns systems during the fast decay from the plateau in both ZTF bands, stressing its different behavior compared to the other systems.

\subsubsection{Mass-transfer Rate and Luminosity Estimates}

Since no distance is known to ASASSN-21au, we are unable to determine the accretion rate to compare it directly to expectations from the DIM. However, rough calculations can be made in order to estimate that parameter. Considering the model by \cite{2006Bil} for an AM~CVn of period 58 min and a massive ($\sim 1 M_{\odot}$) WD, which seems to be appropriate \citep[considering the case of
SDSS~1137 with P$_{\rm orb}\sim60$ min and recent estimates of masses in AM~CVns;][]{2021vanRoestel_eclipsing}, the absolute $V$ mag of ASASSN-21au should be $\sim$13. 
As there are no $V$ measurements during quiescence, we 
instead use the $g$ value from 
Pan-STARRS ($20.96\pm 0.06$ mags, very similar to Gaia $G$). 
This allows us to estimate
a distance of 400 pc (consistent with our  estimate in \S 3.1, based on peak $L_X$).  
By scaling the luminosities to this value one obtains the relations $L_{X,\,outburst} =1.4\times10^{31} d_{400}^2$~erg/s, during outburst and $L_{X,\,post} =7.3\times10^{30} d_{400}^2$~erg/s
during the early post-outburst cooling phase, where $d_{400}$ is $d / 400$~pc. On the other hand, the NUV ($UVW2$, $UVM2$, $UVW1$) luminosity is
$L_{NUV,\,outburst}=1.6\times10^{34}d_{400}^2$~erg/s and $7.4\times10^{31}d_{400}^2$~erg/s, respectively in each state.
Assuming that the NUV luminosity during outburst is due to accretion on the WD with R $= 0.008$~$R_{\odot}$ (similar to Sirius B, and appropriate for a 1~M$_\odot$ WD) we obtain $\dot M_{NUV,\,outburst}=6.7\times10^{16} R_{008} d_{400}^2 M_1^{-1}$ g/s, where $R_{008}$ is $R / 0.008 R_{\odot}$ and $M_1$ is the mass of the accreting WD in solar masses. The value of $\dot M$ would be even larger if we consider the bolometric luminosity. In fact if the extreme UV luminosity is larger than the NUV one (as expected), $\dot M$ would be larger than $\dot M_{\rm crit}^+(R_{\rm out})=1.2\times10^{17}$ g/s \citep[considering the expressions given in A2 of][and assuming R$_{\rm out,10}$=1.3]{Kotko2012}, as would occur if the full disk is in the hot state. An upper limit on the accretion rate in full quiescence can be determined from the X-ray luminosity during the post-outburst cooling phase\footnote{In full quiescence the X-ray luminosity must be lower than the observed one during the post-outburst cooling phase.}.
We use the X-ray flux for the upper limit, since in the cooling 
phase most of the UV flux presumably originates from the hot WD with some contribution from the illuminated inner accretion disk,
while the accretion flow is thought to radiate optically thin bremsstrahlung, placing most of the output radiative flux into the X-rays.
 We find, using the same 
 parameters, that $\dot M_{qui}$ is  $<3\times10^{13}R_{\rm in,008} d_{400}^2M_1^{-1}$~g/s, which is of the same order as $\dot M_{\rm crit}^-(R_{\rm in})$ if $R_{\rm in} > d_{400}^{1.2}M_1^{-0.08}\times10^{9}$~cm, again as expected if the DIM holds\footnote{Note that contrary to R$_{\rm out}$, the value of R$_{\rm in}$ is not fixed by the orbital parameters.}.

\subsection{The Spectral Energy Distribution of ASASSN-21au in Quiescence}

\begin{figure}
	\includegraphics[width=1.0\columnwidth, trim=0.5cm 6cm 0cm 6cm]{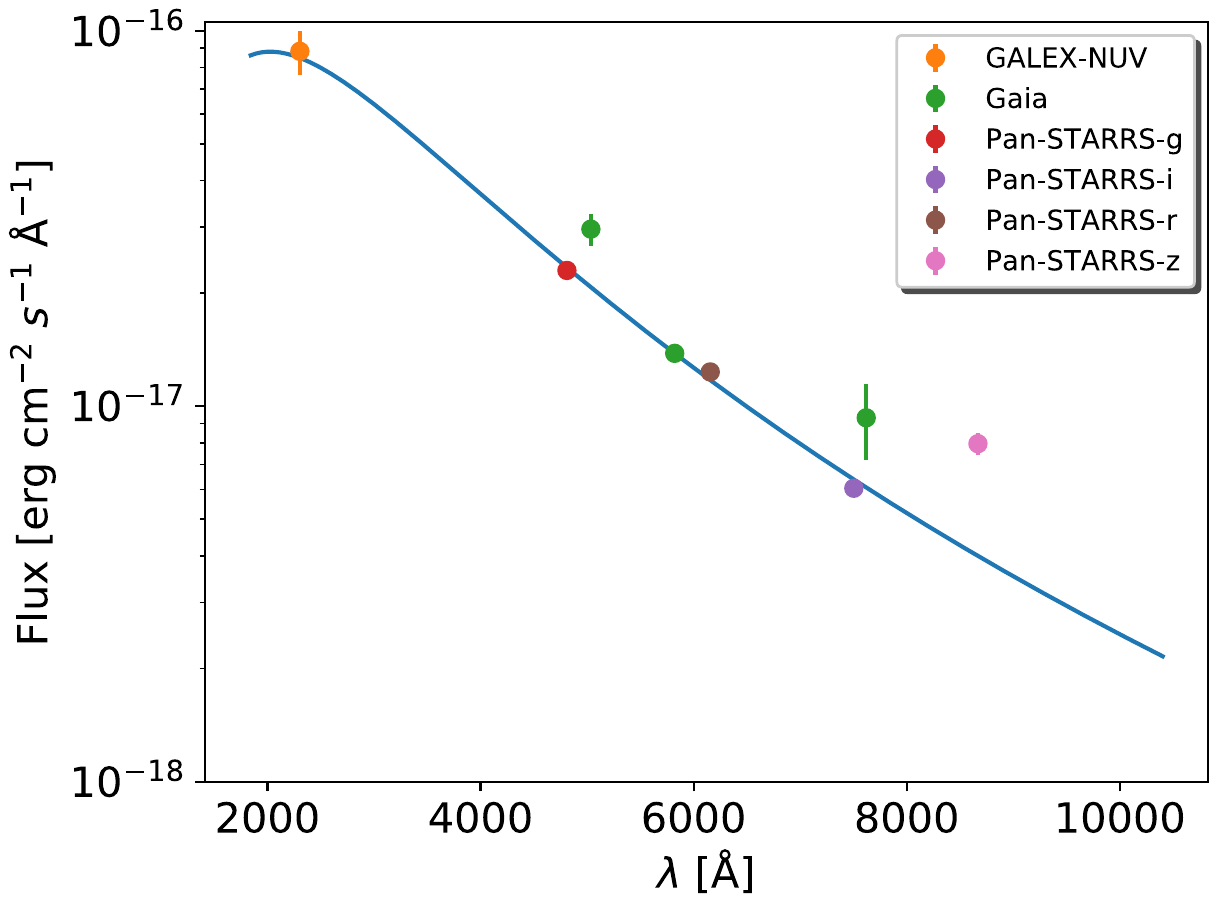}
	\caption{Spectral energy distribution of ASASSN-21au during quiescence using archival data. The blue line denotes the best blackbody fit with a temperature of $14300\pm1600$~K. The z-band point is at $7\sigma$ above the fit, possibly due to an excess. Additional IR data is required to confirm this. }
	\label{fig:ZTF_SED}
\end{figure}

Using archival data obtained during full quiescence we have carried out a spectral energy distribution analysis fitting a blackbody to constrain the temperature of ASASSN-21au. We used the GALEX, Pan-STARRS and Gaia DR3 data, together with a E(g-r) $= 0.05$ \citep[from the 3D dust map of ][]{2019Green_dustmap} and obtained T$= 14300 \pm 1600$~K, which is roughly consistent with expectations from \cite{2006Bil} for an AM~CVn of P$_{\rm orb}\sim58$ min and  $M \sim$1 M$_{\odot}$ WD.
It is possible that there is an IR excess as the Pan-STARRS-$z$ data point is $7\sigma$
above the blackbody fit line (Figure \ref{fig:ZTF_SED}). However,
given that we have only one IR data point, a fit with 2 blackbodies (accreting WD and disk/donor) is very poorly constrained leading to practically meaningless results.
New, deeper 
IR observations are needed to confirm this excess. As showed in \cite{2006Bil}, the contribution of the accretion disk in quiescence should
be minimal in optical, and the hot WD should dominate the optical light in quiescence.

Note that a comparison of the WD temperature based on the spectral energy distribution during the post-outburst cooling phase is not really possible considering the small amount of data obtained in this period. Also, during that phase a contribution from the accretion disk is still expected which would be difficult to disentangle from the WD emission, especially considering the limited data points available. 

\begin{center}
\begin{table*}
\hskip-3.cm
{\footnotesize
\begin{tabular}{|c|c|ccc|c|cc|}
\hline
 & 
\multicolumn{4}{c}{n-Plateau's decay (mag/d)}&
\multicolumn{3}{c|}{n-Cooling phase (mag/d)} \\
&\multicolumn{4}{c}{\hrulefill{}}&\multicolumn{3}{c|}{\hrulefill{}} \\
ID & Dates & ZTF $g$ & ZTF $r$ & Clear & Dates &  ZTF $g$ & ZTF $r$ \\
 & (JD) &($\times10^{-3}$) & ($\times10^{-3}$)& ($\times10^{-3}$) & (JD)& ($\times10^{-3}$) & ($\times10^{-3}$) \\
 \hline
 ASASSN-21au & $2459267-2459269$& $1221.83\pm9.11$ & $1091.67\pm6.67$ 
& $1141.83\pm2.56$ &
$2459285-2459330$   &
$8.11\pm0.50$ & $8.80\pm0.55$ \\
\hline
SDSS~0807 & $2458591-2458816$ & $9.43\pm0.18$ & $8.91\pm0.17$ 
& -- &
$2458816-2459330$ &
$0.83\pm0.13$ & 
$1.24\pm0.10$ \\
\hline
SDSS~1137 &$2458225-2458492$& $1.38 \pm 0.07$ & $2.61\pm0.06$ 
& -- &
$2458568-2459330$  &
$0.05\pm 0.02$ & 
$0.19\pm0.02$ \\
\hline
\end{tabular}
\caption{Values of the slope n when fitting a linear model (n$\tau$ + b) to the magnitude evolution during the decay from the plateau and the post-outburst cooling phase of ASASSN-21au, SDSS~0807 and SDSS~1137.}
\label{decays}
}
\end{table*}
\end{center}

\subsection{The Varying Color Evolution of ASASSN-21au: the Binary Reveals a Blue Superoutburst}
\label{color}

Recently, it has been shown that outbursts in long-period AM~CVns have a color evolution that is not compatible with the one expected by 
the DIM \citep{2020RS08,2021RS}. In SDSS~0807 and SDSS~1137 the AM~CVns become redder and brighter as they reach the maximum of the outburst. That color evolution together with the duration and amplitude of the outbursts suggest that  
the outbursts are confined to the outer parts of the disk, perhaps due to enhanced mass-transfer from the donor. 

In Figure \ref{fig:colorZTF} we present the color evolution of ASASSN-21au using the data from ZTF $g$ and $r$ before $JD*=349$ (when the precursor started). We see that the binary becomes slightly redder compared to quiescence as it increases its brightness, reaching ZTF $g-r>0.5$. Note that measurements have large error bars because the binary is faint and is close to the limiting magnitude of ZTF. However, despite the large scatter, one finds that 
the color behavior of ASASSN-21au is completely compatible with the color evolution observed in SDSS~0807 and SDSS~1137, suggesting then that such part of the light curve in ASASSN-21au can also be explained under an enhanced mass-transfer scenario. Note that during quiescence the binary is blue because the WD is dominating the emission. 

\begin{figure}
	\includegraphics[width=1.0\columnwidth, trim=0.5cm 0cm 0cm 0.5cm]{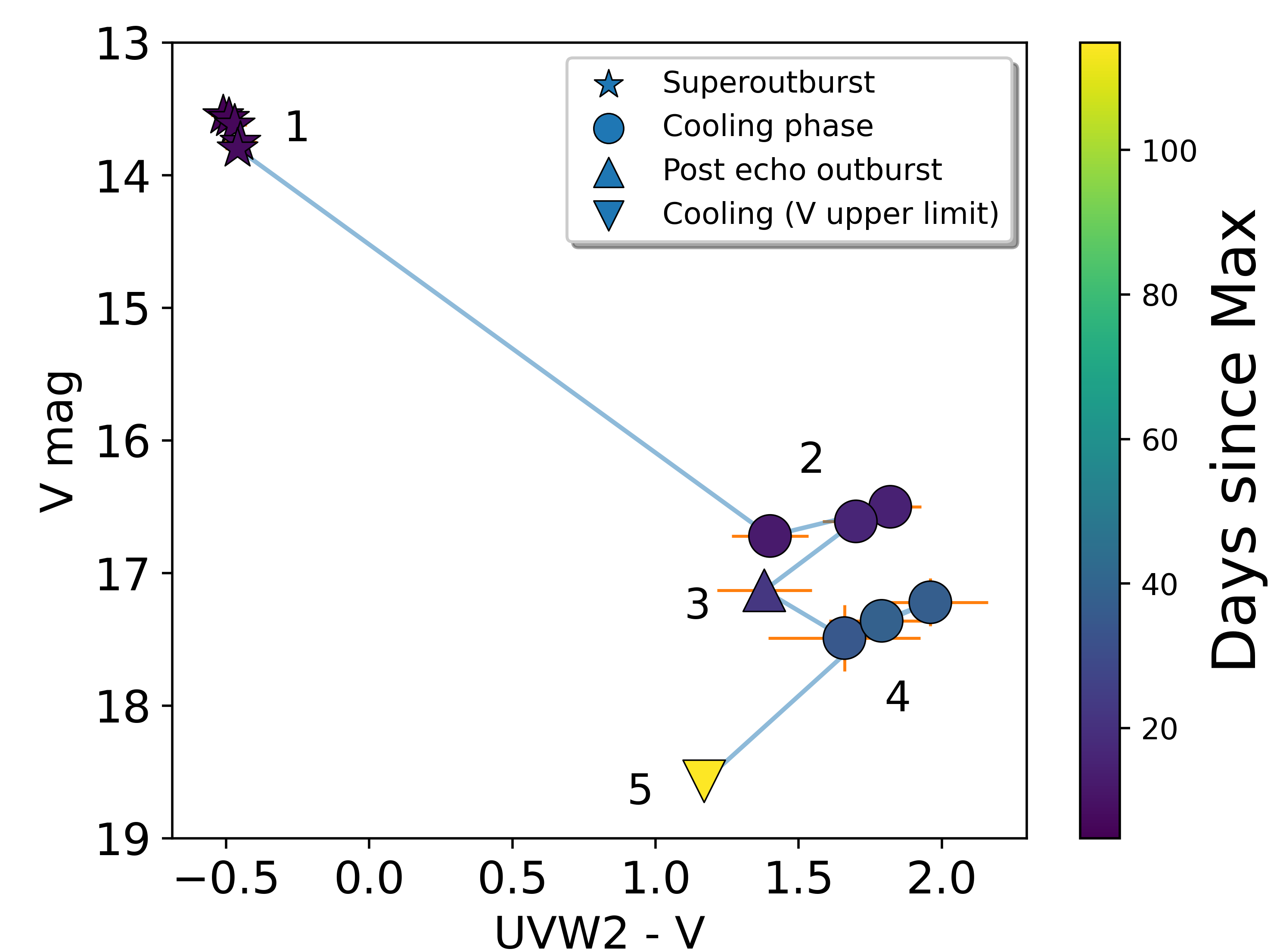}
    \caption{Color evolution for the AM~CVn ASASSN-21au with an orbital period $\sim58$~min during superouburst as indicated by the UVOT data in the NUV band $UVW2$ and the optical filter $V$. The dominant sources of emission are marked with numbers in the plot and are: 1,3 = Disk + WD + boundary layer, 2,4 = Disk, 5 = WD. The point at UVW2--V$= 1.38$ is bluer than the preceding ones because it was taken at the end of the echo outburst. Afterwards the binary 
    reddens again. The general color evolution of ASASSN-21au is compatible with expectations from the DIM \citep{2020Hameury}. However, the pattern is opposite to the one followed by 
    SDSS~0807 and SDSS~1137 \citep{2021RS} indicating that the mechanism   
    that drives the outbursts in these long-period systems is different.
    }
    \label{fig:color_mag}
\end{figure}

On the other hand, in Figure \ref{fig:color_mag} we present the color evolution of ASASSN-21au using the simultaneous UV and optical data obtained by \emph{Swift} during superoutburst and post-outburst cooling phase. We chose the bluest ($UVW2$) and reddest ($V$) UVOT filters to trace the behavior.  
Figure \ref{fig:color_mag} shows 
that ASASSN-21au follows an opposite pattern to that observed in SDSS~0807 and SDSS~1137. In the case of ASASSN-21au, the binary becomes bluer and brighter as it is closer to the peak, and as it cools down it becomes redder. Note that in Figure \ref{fig:color_mag}, there is a second turn towards bluer colors at a $V$ magnitude of $17.1$, which is due to a measurement obtained at the end of the echo outburst, when the accretion disk and accreting WD were still hot. After that, the binary continues becoming redder and fainter. The last blue turn in that diagram corresponds to the accreting WD becoming the dominant emitting source, being much brighter in UV than in optical, where just upper limits were detected in $V$. A ZTF color evolution of ASASSN-21au during outburst in $g$ and $r$ is also displayed in Figure \ref{fig:ZTF_color_mag_complete} of the Appendix.

The color evolution of ASASSN-21au during superoutburst is similar to that of the binary systems SDSS~1411 \citep{2021RS}, PTF~0719+4858 and SDSS~J1043+5632 \citep{2021PichardoM}.
It is also compatible with expectations from the DIM \citep{2020Hameury}.
This strengthens our conclusion
that the main mechanism responsible for the superoutbursts (after $JD*=349$) in ASASSN-21au is a disk instability.

\section{Discussion}

From the theoretical point of view, two parameters that play a very important role in the DIM are: the mass-transfer rate, and the truncation radius of the inner disk. The disk is unstable if the mass-transfer rate $\dot M$ is in the range $\dot M_{\rm crit}^- (R_{\rm in}) < \dot M < \dot M_{\rm crit}^+ (R_{\rm out})$ where $\dot M_{\rm crit}^-$ and $\dot M_{\rm crit}^+$ are the critical mass-transfer rates for being on the cold and hot branches, estimated at the inner and outer disk radii, respectively. In the case of CVs, the actual mass-transfer rate is observed to vary significantly between systems at a given orbital period, meaning that the actual mass-transfer rate deviates from its secular mean which, in principle, depends mainly on the orbital parameters \citep[e.g.][]{2011Knigge, 2018Dubus}. It is then not difficult to believe that something similar occurs in some AM~CVns.

Furthermore, disk truncation in CVs has also been demonstrated to exist. 
For example, observational estimates of the
inner disk radius during quiescence have
been reported to be larger than the WD radius \citep{2012Balman}. 
Disk truncation occurs if the WD's 
magnetic field is strong enough to alter the accretion flow,
or if the accretion flow becomes optically thin and geometrically thick close to the WD. There are no strong observational constraints on this, but there is a mere indication that the latter mechanism might be preferred in some systems \citep[see e.g.][]{2021hameury}. Disk truncation is important for cold and stable accretion disks in X-ray binaries and CVs \citep[e.g.][]{2018Dubus}. But in AM~CVns that is not a necessary condition for the binary to have a cold and stable disk, because extremely small mass-transfer rates are expected to be present in long-period AM~CVns due to their evolution \citep[e.g.][]{2006Bil, 2007Deloye} which would be sufficient to meet the condition $\dot M_{\rm crit}^- (R_{\rm in}) > \dot M$ and so to have cold and stable disks \citep[e.g.][]{Kotko2012}. It is also important to point out that these extremely low mass-transfer rates in AM~CVns could make the accretion disks somehow different from those in CVs. For example, the optically thick assumption might not be valid. Furthermore, an additional parameter to the previously mentioned ones could be at play in AM~CVns: the metallicity, which ultimately translates in the kind of donor, and which would affect the optical thickness of the disk as well. \looseness=-10

The characteristics of the light curve of ASASSN-21au, its similarities and differences to the light curves of SDSS~0807 and SDSS~1137 are then compatible with the following scenario: enhanced mass-transfer could be the mechanism that triggers the brightness increases in the 3 AM~CVns. Thus explaining their low amplitude, slow brightness increase and red color evolution. The red color is in agreement with the outer parts of the disks being the dominant sources of emission, perhaps due to a larger contribution from the hot spot. 

It is important to note that the long outbursts observed in SDSS~0807 and SDSS~1137 appear to be rather unique in the zoo of compact binaries. (i) Even if reports of long term increases before a DIM outburst have been reported in other binaries, they are quite different. For example, CVs of the type Z~Cam can show long standstills between (short) normal
outbursts due to an increase in the accretion rate, but the behavior and timescales (e.g. much faster rise and decay) of a Z~Cam star are different to the ones observed in SDSS~1137 and SDSS~0807 \citep[see e.g.][for a long term light curve of Z Cam]{1998Oppenheimer}. Low mass X-ray binaries have also shown years duration standstills between outbursts. However, such mechanism is also analogous to the one of Z~Cam stars \citep[e.g. Swift J1753.5-0127,][]{2019Shaw}. (ii) SDSS~1137 and SDSS~0807 show clearly distinguishable outbursts instead of erratic transitions between high and low states as observed in other CVs. These include systems with high mass-transfer rates, such as the VY Scl and some intermediate polars as FO~Aqr \citep[see e.g.][]{2020Kennedy} and AM~Her (which shows low states). These contrasts should, however, not come as a surprise, given the  large difference in the secondaries of AM~CVns and CVs. It is then clear that the outbursts in the long-period AM~CVns SDSS~1137 and SDSS~0807 do not have an origin in disk instabilities and so far have no counterpart on CVs nor X-ray binaries. 

If there are mass enhancements for ASASSN-21au, they could even occur at different rates, considering the 2 different brightness levels observed prior to the superoutburst.
At a later stage, mass enhancement could have produced disk instabilities in ASASSN-21au, either due to a further increase in the mass-transfer rate value or to the accretion disk accumulating enough mass at the same increased mass-transfer rate before the superoutburst. 
This would explain the fast, large amplitude and blue superoutburst observed in ASASSN-21au which is consistent with expectations from the DIM. The presence of mass enhancement would also explain the increased magnitude after the superoutburst in ASASSN-21au and the presence of an echo outburst at later times (see \S \ref{light_curve_sec}). 
The observations of ASASSN-21AU during superoutburst clearly contrast to what has been observed in other AM~CVn systems with similar long orbital periods, thus revealing a dichotomy in their behavior.

It is unclear why instabilities were not observed in SDSS~0807\footnote{Given the slightly blue color evolution of SDSS~0807 near the peak of the outburst, \cite{2021RS} discussed the possibility that disk instabilities could have been developed during that phase. However, despite that scenario cannot be discarded, it is not very likely considering the observed values of g-r in that phase and the timescales expected from the DIM as discussed in this paper, which were not observed in the case of SDSS~0807.} and SDSS~1137 but under an enhanced mass-transfer scenario, the mass-transfer rate could have been low enough such that the condition $\dot M_{\rm crit}^- (r_{\rm in}) > \dot M$ was never violated. As mentioned before, brightness increases have been observed to exist previous to DIM outbursts in other accreting WDs (e.g. SS~Cyg) and X-ray binaries \citep[e.g.][]{2016Bernardini, 2020Goodwin, 2021Kimura}. However, we point out that one should be cautious when comparing such phenomena to the event presented in this manuscript given the difference in mass-transfer rates, as well as in the secondary's structure and composition. For example, SS~Cyg has a mass-transfer rate in quiescence which is $\sim4$ orders of magnitude larger \citep{2013MillerJones} than that expected for ASASSN-21au, SDSS~0807 or SDSS~1137, which for CVs, places SS~Cyg well within the expected unstable regime of the DIM, producing then consistent results with that model. On the other hand, AM~CVns systems with long orbital periods must have a mass-transfer rate higher than the secular mean for a disk instability to be triggered, and, as discussed previously, when the value of the mass transfer rate is extremely low, it can lead to different disk behavior. 

\section{Conclusion}

In this paper we have shown observational evidence that a dichotomy exists in the outburst properties of long-period AM~CVns.
Short outbursts are most likely caused by a disk instability, whereas outbursts lasting for a year or more cannot be explained by the DIM and are probably due to a long lasting mass-transfer event. The whole behavior of  ASASSN-21au is consistent with a scenario in which initial mass enhancement produced disk instabilities at later times, contrary to what has been observed in other long-period AM~CVns. Additional studies of similar systems are needed to fully confirm this scenario. 
At present, the origin and characteristics of mass-transfer events remain a mystery; they have also been postulated in systems such as CVs close to the minimum period that exhibit rare superoutbursts (the WZ Sge systems). The causes of such phenomena need to be further investigated through high quality and multiwavelength observations of AM~CVns both during outbursts and in quiescence.

\section*{Acknowledgements}

We thank the referee for her/his comments which improved the manuscript. LERS was supported by an Avadh Bhatia Fellowship at the University of Alberta and a Gruber-IAU Fellowship during the realization of this work. CH is supported by NSERC Discovery Grant RGPIN-2016-04602. The authors acknowledge the \textit{Swift} team for scheduling the target of opportunity requests. The authors also acknowledge the ATLAS, ASAS-SN, Pan-STARRS and VizieR data bases for providing part of the data presented in this manuscript. 
This work has made use of data from the European Space Agency (ESA) mission
{\it Gaia} (\url{https://www.cosmos.esa.int/gaia}), processed by the {\it Gaia}
Data Processing and Analysis Consortium (DPAC,
\url{https://www.cosmos.esa.int/web/gaia/dpac/consortium}). Funding for the DPAC
has been provided by national institutions, in particular the institutions
participating in the {\it Gaia} Multilateral Agreement. We acknowledge the Photometric Science Alerts Team (\url{http://gsaweb.ast.cam.ac.uk/alerts}). Based on observations obtained with the Samuel Oschin 48-inch Telescope at the Palomar Observatory as part of the Zwicky Transient Facility project. ZTF is supported by the National Science Foundation under Grant No. AST-1440341 and a collaboration including Caltech, IPAC, the Weizmann Institute for Science, the Oskar Klein Center at Stockholm University, the University of Maryland, the University of Washington, Deutsches Elektronen-Synchrotron and Humboldt University, Los Alamos National Laboratories, the TANGO Consortium of Taiwan, the University of Wisconsin at Milwaukee, and Lawrence Berkeley National Laboratories. Operations are conducted by COO, IPAC, and UW.

\clearpage

\appendix

\section{The color evolution of ASASSN-21au during superoutburst}

\begin{figure}[!h]
	\includegraphics[width=1.0\columnwidth, trim=0.5cm 0cm 0cm 0.5cm]{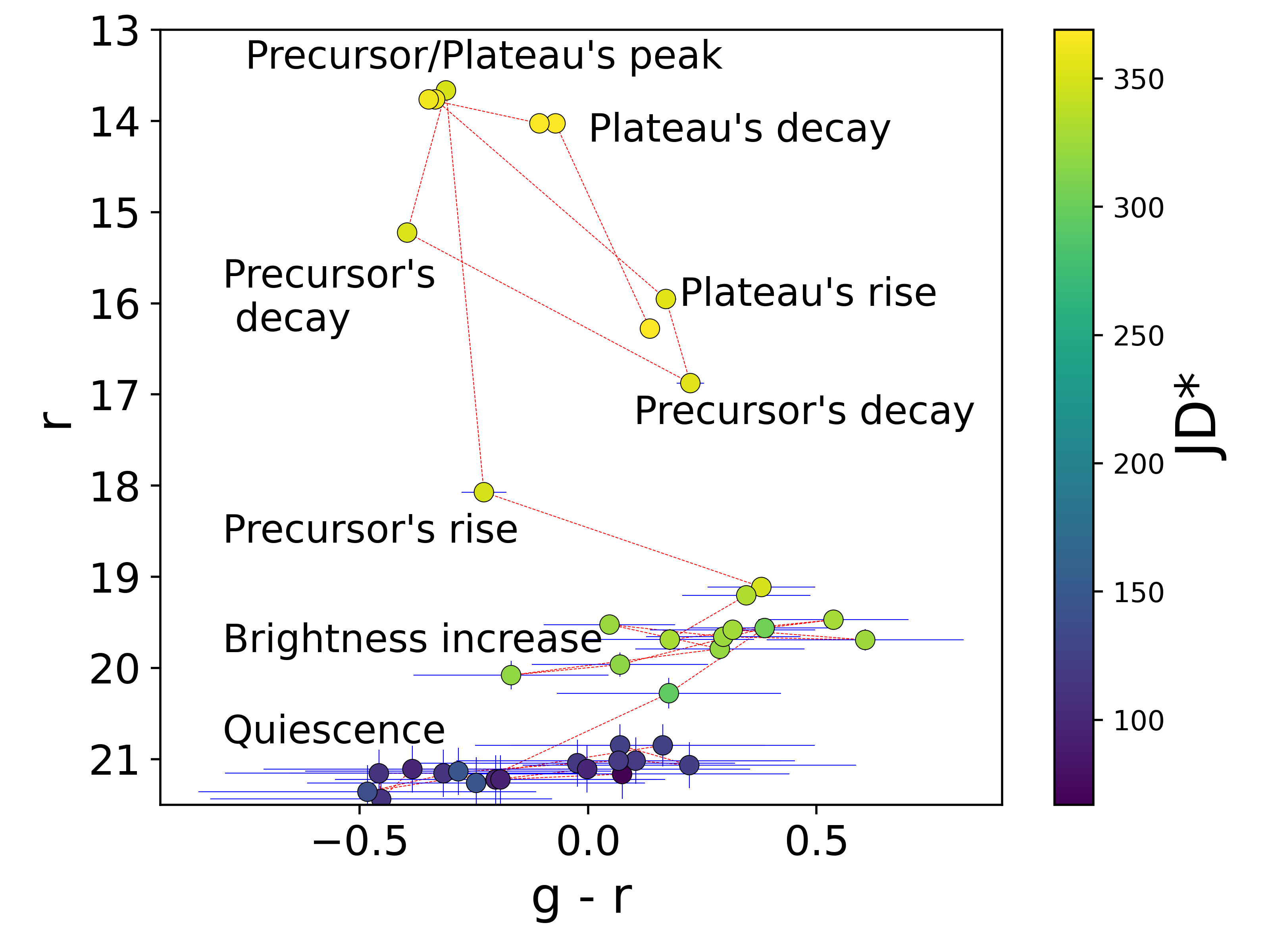}
	\caption{Color evolution based on ZTF data of ASASSN-21au with an orbital period $\sim58$~min up to the end of the plateau phase of the superoutburst. The binary is initially blue due to the WD dominating the emission during quiescence. During the increased brightness levels 1 and 2 (see \ref{light_curve_sec}) the binary increases its brightness and it becomes redder. When the precursor of the superoutburst starts, ASASSN-21au becomes blue. The same behavior is observed when the rise of the plateau occurs. The binary becomes redder during the decay of the precursor and the decay of the plateau.}
	\label{fig:ZTF_color_mag_complete}
\end{figure}

Figure \ref{fig:ZTF_color_mag_complete} shows the color evolution of ASASSN-21au using the ZTF data.

\clearpage
\bibliography{biblio}{}
\bibliographystyle{aasjournal}



\end{document}